# How do intrinsically disordered protein regions encode a driving force for liquid-liquid phase separation?

Wade Borcherds[1], Anne Bremer[1], Madeleine B. Borgia[1], Tanja Mittag[1]*

[1] Department of Structural Biology, St. Jude Children's Research Hospital, Memphis, TN, USA

* Correspondence: tanja.mittag@stjude.org

**Highlights**

- Not all IDRs undergo liquid-liquid phase separation.
- IDRs are not necessary for protein phase separation.
- A stickers-and-spacers framework helps conceptualizing IDR phase behavior.
- The valence and patterning of stickers determine IDR phase behavior.
- Spacers determine the interplay of percolation and phase separation.

**Graphical abstract**

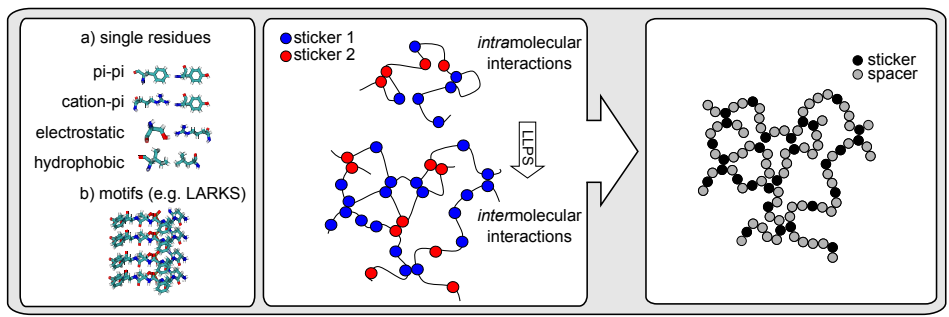

**Abstract**

Liquid-liquid phase separation is the mechanism underlying the formation of biomolecular condensates. Disordered protein regions often drive phase separation, but molecular interactions of disordered protein regions are not well understood, sometimes leading to the conflation that all disordered protein regions drive phase separation. Given the critical role of phase separation in many cellular processes, and that dysfunction of phase separation can lead to debilitating diseases, it is important that we understand the interactions and sequence properties underlying phase behavior. A conceptual framework that divides IDRs into interacting and solvating regions has proven particularly useful, and analytical instantiations and coarse-grained models can test our understanding of the driving forces against experimental phase behavior. Validated simulation paradigms enable the exploration of sequence space to help our understanding of how disordered protein regions can encode phase behavior, which IDRs may mediate phase separation in cells, and which IDRs are in contrast highly soluble.

**Introduction**
Biomolecular condensates are non-stoichiometric assemblies of biomolecules that lack a surrounding lipid membrane and are responsible for extensively compartmentalizing eukaryotic cells [1,2]. Biomolecular condensates are formed via liquid-liquid phase separation (LLPS), a process that occurs when the concentration of macromolecules exceeds a threshold concentration (i.e., the so-called saturation concentration ($c_{sat}$)). Under these conditions, a density transition occurs [3,4] that results in the formation of a dense phase, in which the macromolecules are enriched, and a dilute phase, which is relatively depleted of the macromolecules. The dense phase often appears as dense liquid-like droplets in a surrounding dilute phase (Figure 1).

LLPS is now understood to play ubiquitous roles in fundamental cellular processes, including key roles in promoting the assembly of the mitotic spindle during cell division [5], the cellular stress response [6,7], transcription [8-12], and RNA metabolism [13]. Alteration of LLPS behavior has been linked to several diseases with examples including amyotrophic lateral sclerosis (ALS) [14,15] and frontotemporal dementia [16], and Alzheimer's disease [17].

While many biological processes seem to involve phase separation, the question has been raised as to whether phase separation per se is required for a given function. Instead, small complexes may be the functional species and phase separation a byproduct of the crowded environment in cells. How might one test whether phase separation is absolutely required for a specific function or the critical factor in a disease process? If one could introduce protein variants with a wide spectrum of phase behaviors, and then compare their phase behavior to the strength of the related phenotype, a high correlation should indicate that phase separation is important for the process. This goal requires the quantitation of protein phase behavior. A quantitative understanding of the driving forces for phase separation will thus be useful and would also enable the engineering of stimulus responsive materials, e.g., for advanced drug delivery. The development of several neurodegenerative diseases seems to be directly linked to phase separation, and mutations driving pathogenesis may do so by altering the phase behavior of the mutated proteins [16]. Being able to predict the effect of mutations on the driving force for phase separation would help in classifying mutations as benign or pathogenic. Last, but not least, being able to predict from the sequence whether a protein will undergo phase separation, and how strong its driving force to do so is, demonstrates conceptual (or even quantitative) understanding of the underlying interactions. Given the dependence of the driving force for phase separation on the concentration of small molecules, ions, binding partners, and on crowding, this is all but a simple challenge. Uncovering the physical principles underlying biomolecular phase separation and quantifying contributions of other components step by step will help us to arrive at the ability to predict context-dependent phase behavior from the protein sequence.

**Multivalent interactions drive phase separation**
Seminal work by Mike Rosen and coworkers demonstrated that physiological phase separation of proteins and/or RNA is mediated by multivalent interactions [18]. In the classical systems they studied, pairs of proteins underwent phase separation above threshold concentrations and at suitable molar ratios. In these systems, protein A encoded tandem repeats of modular binding domains connected by flexible linkers, and protein B was a disordered protein with multiple short linear motifs, each of which could bind to the modular domains in protein A (Fig. 1). If the repeat domains in protein A were RNA-binding domains, molecule B was an RNA molecule. These multivalent interactions enable the formation of three-dimensional networks of molecules, which span the volume of the resulting dense-phase droplets. Proteins can also achieve multivalence through discrete oligomerization as in the case of NPM1 [19], or through indefinite linear oligomerization as in the cases of SPOP [20] and TDP-43 [13].

On the other hand, numerous intrinsically disordered proteins, or intrinsically disordered protein regions (IDRs), can undergo phase separation via homotypic interactions, i.e., without binding partners (Fig. 1). What are the driving forces that mediate phase separation of IDRs? And is the ability to undergo phase separation a characteristic property of IDRs as many recent papers seem to hypothesize? In this review, we will make the case that only a subset of IDRs has a strong enough driving force to undergo phase separation that is physiologically relevant. Other IDRs are highly soluble. IDRs with a high fraction of charged and/or proline residues, which promote solvation rather than attractive intrachain interactions [21,22], are not likely to mediate phase separation. We will lay out that phase separation of IDRs and domain-motif systems can be understood within the same conceptual framework. We will discuss the role of adhesive motifs (or so-called "stickers") and their connecting "spacers", and how sticker valence, pairwise interaction strength and patterning shape the phase behavior of IDRs.

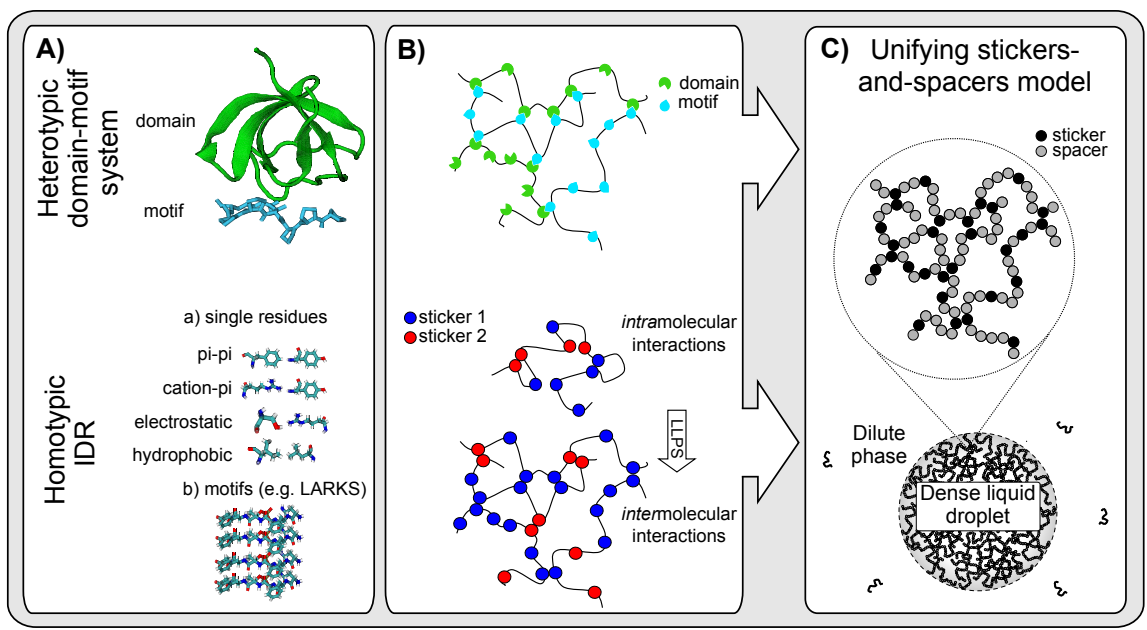

**Figure 1. Conceptualizing liquid-liquid phase separation of IDRs.** The interactions that drive LLPS in domain-motif systems and IDRs can both be described by the stickers-and-spacers framework. Stickers are adhesive elements that contribute to the main interaction potential, and they are connected by largely non-interacting spacers. **(A)** Heterotypic LLPS in domain-motif systems, e.g., between a folded SH3 domain and a proline-rich motif (PRM) (top, PDB ID: 1SEM). LLPS of IDRs can be mediated by a multitude of multivalent interactions. These may include interactions of individual residues or longer motifs, e.g., LARKS (bottom, PDB ID: 6CF4). **(B)** SH3 tandem repeats connected by linker regions can phase separate in the presence of tandem repeats of PRMs (top). The homotypic *inter*molecular interactions that drive phase separation of IDRs are satisfied *intra*molecularly in the dilute phase (bottom). **(C)** In the stickers-and-spacers framework, SH3 domains and PRMs are stickers, and the connecting linkers are spacers. For IDRs, the single residues or motifs are the stickers and the intervening residues spacers.

**Phase separation of IDRs**

Many IDRs are sufficient for mediating phase separation, for example the IDRs of FET family proteins, members of the heterogeneous nuclear ribonucleoprotein family, including hnRNPA1 [23], hnRNPA2B1 [24], RNA helicases Ddx4 [25] and Laf-1 [26] among others. Also, biomolecular condensates contain many proteins with IDRs [27] suggesting a role in protein recruitment and perhaps shaping condensate properties. Different biomolecular condensates seem to enrich IDRs with specific sequence features; nucleoli contain arginine-rich IDRs, while nuclear speckles contain serine/arginine-rich IDRs. Stress granules contain many IDRs enriched in small polar residues and a small fraction of aromatic and charged residues, often called prion-like domains [28,29]. The distinct sequence features in different biomolecular condensates suggest that IDRs can encode phase behavior in a number of different ways, and that IDRs may confer specificity in phase separation, controlling which proteins co-phase separate. Many of these IDR sequences comprise a limited set of amino acid types inspiring their designation as low complexity domains (LCDs) [30,31]. It is important to note that low sequence complexity is neither required nor sufficient for an IDR to undergo phase separation.

Common characteristics of IDRs are well suited for a role in phase separation. IDRs lack stable secondary and tertiary structure and are instead characterized by an ensemble of interconverting conformations. Consequently, IDRs access a broader range of conformational space, which may permit a multitude of concurrent interactions that enable the formation of three-dimensional networks of protein molecules. IDRs often harbor many post-translation modification (PTM) sites because they are accessible for modifying enzymes, offering a mechanism for biology to regulate phase behavior [32-34]. They may also be able to form less dense networks compared with those formed by globular proteins. The resulting porous meshwork may be able to recruit other constituents easily and enable enzymatic reactions and signaling events. While these points highlight the important roles protein disorder can play in the functioning of biomolecular condensates, it is important to note that IDRs are not always drivers of phase separation, and can, in fact, inhibit LLPS [7].

Numerous types of potential molecular interactions have been shown to contribute to the driving force for phase separation of IDRs (Fig. 1B). Additional interactions may form within dense phases due to the high local concentrations of protein [35]. These include hydrophobic [36], electrostatic [25,37], pi-pi [4,38,39], cation-pi [4] and hydrogen bond interactions [40]. Details on the specifics of these interactions have recently been reviewed elsewhere [41,42]. Whether these interactions are mediated through the sidechains of residues in otherwise disordered chains, or whether they involve transient structuring of the backbone and hydrogen bonding is still an open question. The latter possibility has support through the discovery that IDRs may contain repeats of short linear interaction motifs, called low-complexity aromatic-rich kinked segments (LARKS) [43], or Reversible Amyloid Cores (RACs) [44,45], which can form cross-beta-type structure (further reviewed in Peran & Mittag 2020 [46]).

Hence, various interactions can contribute to phase separation of IDRs, as we may well expect from the range of physicochemical properties of amino acids. But how do we know which interactions are most important in a given sequence and whether they are collectively sufficient to drive phase separation? This requires a conceptual as well as quantitative understanding of IDR phase separation.

**Conceptualizing driving forces for phase separation**
Multivalent interactions of domain-motif systems support the formation of three-dimensional protein networks and phase separation (Fig. 1). To make complex systems conceptually, theoretically and computationally tractable, Pappu and coworkers introduced the stickers-and-spacers framework for multivalent biomolecules [3] that draws on previous work on associative polymers [47,48]. The stickers-and-spacers model reduces biomolecules to two types of components, "stickers" and "spacers". Stickers are defined as the adhesive elements that contribute the main interaction potential. Spacers connect the stickers and influence their ability to interact with each other and the ability of the biomolecule to interact with solvent. In the case of strongly solvating spacers, density transitions may even be suppressed although percolation gives rise to system-spanning gels at high enough concentrations [3].

The stickers-and-spacers framework for domain-motif systems can be transferred to IDRs. In this case, all stickers are on a single chain and *intra*molecular interactions are swapped for *inter*molecular interactions above threshold concentrations (Fig. 1B). This may not seem obvious initially because we understand determinants of affinity and specificity in domain-motif interactions based on readily available structural information, which is missing in the case of IDR-IDR interactions. Neither the structural features of these interactions nor their affinities are well understood, and it remains to be determined how much specificity they can encode. But indeed, as in domain-motif systems, the interaction strengths between stickers, their valence, their patterning in the sequence, and the properties of the spacers determine the driving forces for phase separation [38,49] and we will make the case for this below.

**Applying the stickers-and-spacers framework to IDRs**
To apply the stickers-and-spacers framework to the understanding of IDR phase separation obviously requires the identification of the stickers and the spacers. This has been accomplished in two ways. In the first strategy, individual residue types or short motifs in the sequence are mutated and the effect on the driving force for phase separation is determined [4,7,39]. If the removal of a residue type or motif increases the saturation concentration, it is identified as a sticker. The contributions of tyrosine, phenylalanine, arginine and lysine residues have been evaluated in this manner in the FET family of RNA-binding proteins, and tyrosine and arginine were identified as stickers [4]. Given that spacer properties also influence phase behavior (see below), this approach may not only identify stickers.

The second strategy takes advantage of the fact that the same stickers that drive phase separation via *inter*molecular interactions must also mediate *intra*molecular interactions in the dilute regime. Recent work by Mittag, Pappu and coworkers identified phenylalanine and tyrosine residues as the residue types that formed the most frequent contacts along the chain of the prion-like LCD of hnRNPA1. The aromatic residues were identified by a combination of NMR spectroscopy and Monte Carlo simulations, and their ability to form cohesive interactions was demonstrated via their ability to compact the chain in a manner that depended on the number of aromatic residues [38]. They then demonstrated that the aromatic residues were indeed the main drivers of LLPS. Titrating the aromatic content in a set of sequence variants resulted in predictable changes in phase behavior.

To implement the stickers-and-spacers model in simulations or via an analytical theory, the interaction strengths of the stickers need to be determined. In the case of hnRNPA1, the interaction strength of pairwise aromatic-aromatic interactions could be extracted from the global dimensions of the set of variants in which the number of aromatic residues was titrated. Small interaction strengths were also assigned to sticker-spacer and spacer-spacer interactions. These parameters were used in on-lattice, coarse-grained simulations in which each amino acid was represented by a single bead, and only two types of beads were used, i.e., stickers and spacers. The simulations with the parameterized stickers-and-spacers model quantitatively predicted the phase behavior of the variants, demonstrating that a simple model containing only the sticker number, position and interaction strength, was able to recapitulate experimental phase behavior. The broader applicability of the model was demonstrated by successfully predicting the phase behavior of the related FET family protein FUS without the need to reparametrize the model [38].

The stickers-and-spacers framework has also been used to implement analytical mean-field theories. The driving force for phase separation should be proportional to the likelihood of forming sticker-sticker interactions. Hence, the saturation concentration ($c_{sat}$) should be inversely proportional to $N^2$, where $N$ is the number of stickers. In the case of FET family proteins, in which tyrosine and arginine residues were identified as stickers that interact with each other, $c_{sat}$ is inversely proportional to the product of the number of tyrosine ($N_Y$) and arginine residues ($N_R$), i.e., to $N_Y*N_R$ [4]. This analytical model can be generalized also to explicitly take into account tyrosine-tyrosine and tyrosine-arginine interactions [4,49,50] or other combinations of sticker interactions.

The stickers-and-spacers framework provides insights into the functional reasons for the sequence degeneracy of LCDs. The relatively uniformly spaced aromatic residues in a background of small polar residues encode repeats of stickers connected by spacers that enable sticker-interactions. The sequence degeneracy therefore encodes the multivalency necessary for phase separation.

The current instantiations of stickers-and-spacers models for IDRs are valid for prion-like LCDs [4,38,50]. To apply the models to other IDRs, one must have *a priori* insight that the stickers are the same as in prion-like LCDs, or the model must be extended to fit the new flavor of IDR. This entails *de novo* identification of stickers and careful parametrization of sticker-sticker, sticker-spacer and spacer-spacer interaction strengths.

Multiple different sticker types can of course also be introduced into simulation-implementations of the stickers-and-spacers model. Pairwise interaction strengths for each sticker type with itself and for the sticker types with each other would need to be parameterized from experimental data. Sticker-spacer interaction strengths might also vary as a function of sticker type. Additional extensions of the stickers-and-spacers model could include effects of different spacer residue types on the solvation of the protein molecule and therefore its phase behavior [3,49]. The on-lattice LASSI model models spacers with positive effective solvation volume as taking up positions on lattice points and spacers with negative effective solvation volume as "phantom" spacers that can overlap [51]. Future development of a model rooted in the stickers-and-spacers framework can be envisioned where a core set of stickers contributing to phase separation have been parameterized on a diverse enough sequence library to account for a majority of LCD sequences, thus eliminating the need for *a priori* knowledge of each sequence's specific stickers. Taken to the extreme, a residue-specific interaction potential for phase separation could be determined from experimental data and implemented in coarse-grained simulations.

**Amino acid-specific interaction potentials**
The idea of a model for IDR phase separation that relies on known characteristics of the amino acids and their interaction potentials has been pursued by Mittal and coworkers. It has been implemented in off-lattice coarse-grained simulations in which each amino acid residue is represented as a bead that models chemical properties like charge and size [52]. They then include an interaction potential for each amino acid type. The full matrix would include the interaction potential between every amino acid to every other amino acid resulting in 210 (20*19/2 + 20) parameters, which is a high level of complexity that is difficult to justify given the problems in parameterizing it properly. Therefore, a single fitted interaction potential is used for each amino acid, as previously parameterized by Kim and Hummer [52]. With these coarse-grained simulations, Mittal and coworkers can predict phase diagrams that are sensitive to sequence variations that perturb the net charge of the protein [53].

**Sequence patterning determines the phase behavior**
The patterning of interacting residues in IDRs also determines their phase behavior as was first shown for the patterning of oppositely charged residues in the IDR of Ddx4 [25]. Mittal and coworkers were able to use the insights from their coarse-grained simulations to develop an analytical framework, and they demonstrated the effect of altering the patterning of the charged residues: increasing the separation of the charged residues increases the driving force for phase separation, while evenly distributing the charged residues reduces it [54]. These observations were consistent with the experimental data on the IDR of Ddx4 and of a variant that changed the charge patterning [25]. Using these simulations they identified a segment in the intrinsically disordered RGG domain of Laf-1 that more strongly contributed to phase separation than any other similarly sized fragment in the sequence, and confirmed their findings experimentally, emphasizing the utility of this framework [54].

The coarse-grained simulations by Mittal and coworkers also revealed patterning of hydrophobic and charged residues as important parameters for phase behavior and single-chain collapse. They have developed a sequence hydropathy and a sequence charge decoration parameter (SHD and SCD, respectively), which quantify the distribution of hydrophobic and charged residues in the sequence and help predict single chain collapse and phase behavior [55].

Hue Sun Chan and coworkers have implemented an on-lattice coarse-grained simulation paradigm that employs the random phase approximation theory to account for the effects of patterning of charged residues in the sequence [56]. In addition, they use a mean-field correction to account for cation-pi interactions [57,58]. With this model they are able to qualitatively predict differences in phase behavior of synthetically designed polypeptide sequences with the same sequence composition but drastically different patterning [59]. They demonstrate the effects charge clustering has on phase behavior. When oppositely charged residues are well mixed, the driving force for phase separation is low. However, when the oppositely charged residues are segregated resulting in clusters of acidic or basic residues, the driving force for phase separation is higher. As the clustering of charged residues increases, the interaction potential of each cluster increases, and the effective valence decreases.

An inherent drawback of analytical models is an insensitivity to conformational preferences. These conformational preferences govern ensemble dimensions which are predictive of phase behavior [34,38,55,60]. Hue Sun Chan and coworkers have recently incorporated sequence-dependent renormalizations to better account for conformational heterogeneity. With this refinement, the theory now predicts realistic phase diagrams that can be reasonably fit to the phase diagrams of the highly charged Ddx4 LCD and a scrambled variant at varying salt concentrations [61].

In the case of the hnRNPA1 LCD, in which aromatic stickers emerged as the major driving force for phase separation, the aromatic residues are far more evenly distributed than expected for a random distribution. This is also the case for many other IDRs, implying evolutionary pressure on the spacing of aromatic residues. This raises the question of how sticker patterning affects phase behavior. It was found that clustering of aromatic residues into several patches in the sequence resulted in the formation of amorphous aggregates instead of liquid condensates. Simulations using the stickers-and-spacers model provided insight into the underlying mechanism of this altered state. The clusters of aromatic residues effectively produced very strong stickers, whose interactions were long lived, giving rise to solid assemblies [38].

**How altering the driving forces affects phase behavior**
Stickers-and-spacers simulations and other coarse-grained simulations can be used to test which sequences may, or may not, phase separate at given concentrations, as well as provide insight into how a given mutation may impact the phase behavior of a protein. These simulations enable the systematic exploration of IDR phase behavior as a function of sticker valence, interaction strength and spacer properties and have resulted in the insights summarized in Figure 2 [3,38,49,62]. Increasing the sticker valence or strength of the interactions will increase the driving force for LLPS, lower the saturation concentration, and result in a wider two-phase regime [63,64] (Fig. 2B and C). The distance between stickers, which is determined by the number of spacer residues and their effective solvation volume, also modulates phase behavior. Extensive separation of the stickers reduces the cooperativity of their contributions and decreases the driving forces for phase separation [3,49]. Increased spacing between stickers can be achieved by a greater number of intervening residues or by highly solvated residues that increase the stiffness and local or overall solubility of the IDR (Fig. 2D and E). On the other hand, decreasing the separation of stickers eventually creates clusters of stickers. Clustering of stickers alters the driving force for phase separation in several ways (Fig. 2F). A cluster may behave like a single sticker with a stronger interaction potential while clustering also reduces the effective valence of the interactions.

In IDRs in which charge-charge interactions contribute to the driving force for phase separation, increased clustering of the charged residues results in an increased driving force for phase separation [25,59,64]. However, increasing the clustering of the stickers can increase the lifetime of the resulting interactions and therefore alter the material properties of the condensate [49], which may ultimately favor irreversible aggregation over LLPS, as found for aromatic clustering in the LCD of hnRNPA1 [38].

**PTMs in the context of the stickers-and-spacers model**
As described above, IDRs are enriched in PTM sites. These modifications can lead to alterations in secondary or tertiary structure and can create or destroy interaction sites. Often, they alter charge and hydrophobicity, thereby altering the molecular interactions which the protein can engage in, promoting or inhibiting the driving forces for LLPS, and modulating the material properties of condensates. Interestingly, whether PTMs enhance or reduce the driving force for phase separation does not only depend on the nature of the PTM but also on the sequence that is modified, i.e., it is strongly context dependent. Examining PTMs in the context of the stickers-and-spacers model helps conceptualize their effects.

Phosphorylation has been shown to play a role in controlling the phase behavior of several proteins as well as affect the sub-compartmentalization within condensates. The addition of the negatively charged phosphoryl group modifies uncharged and polar serine, threonine or tyrosine into negatively charged amino acids. We may expect that tyrosine phosphorylation reduces its interaction strength as an aromatic sticker but enables new charge-charge interactions if an excess of positively charged residues is present in the sequence. Phosphorylation of the spacer residues serine and threonine enhances their effective solvation volumes and thus also reduces phase separation, unless attractive charge-charge interactions are introduced. Phosphorylation of the C-terminal IDR of FMRP increases its phase separation [65]. Conversely, phosphomimetic substitutions in the intrinsically disordered N-terminal prion-like LCD of FUS has been found to reduce phase separation [66]. Finally, the CAPRIN-1/FMRP/RNA system illustrates the complex relationship PTMs can have on phase behavior. Phosphorylation of either protein increases their propensity to co-phase separate compared with their unphosphorylated forms. However, phosphorylation of both reduces their ability to co-phase separate. Furthermore, RNA introduction results in CAPRIN1/RNA subcompartments within the CAPRIN1/pFMRP droplets, but not in the pCAPRIN1/FMRP droplets [67].

Methylation of arginine residues does not involve a change in charge, but alters volume, hydrogen bonding potential, hydrophobicity and charge density. Arginines can be mono-methylated, as well as symmetrically or asymmetrically dimethylated resulting in numerous possible methylation patterns. Preferred sites of methylation are GRG or RGG motifs, which are frequently found in LCDs [25]. Arginine methylation is expected to lower its interaction potential and therefore reduce LLPS propensity [41]. Indeed, arginine methylation has been found to destabilize and dissolve dense phase droplets formed by Ddx4 primarily via electrostatic interactions [25]. Arginine methylation has also been reported to reduce hnRNPA2 phase separation by disrupting arginine-mediated contacts [68], and to disfavor phase separation in the previously mentioned IDR of FMRP by perturbing RGG-mediated RNA-binding interactions [65]. However, the intrinsically disordered C-terminal domain of LSM4 requires symmetric dimethylation for phase separation into P bodies, and mutating arginine residues diminishes P body formation in cells [69].

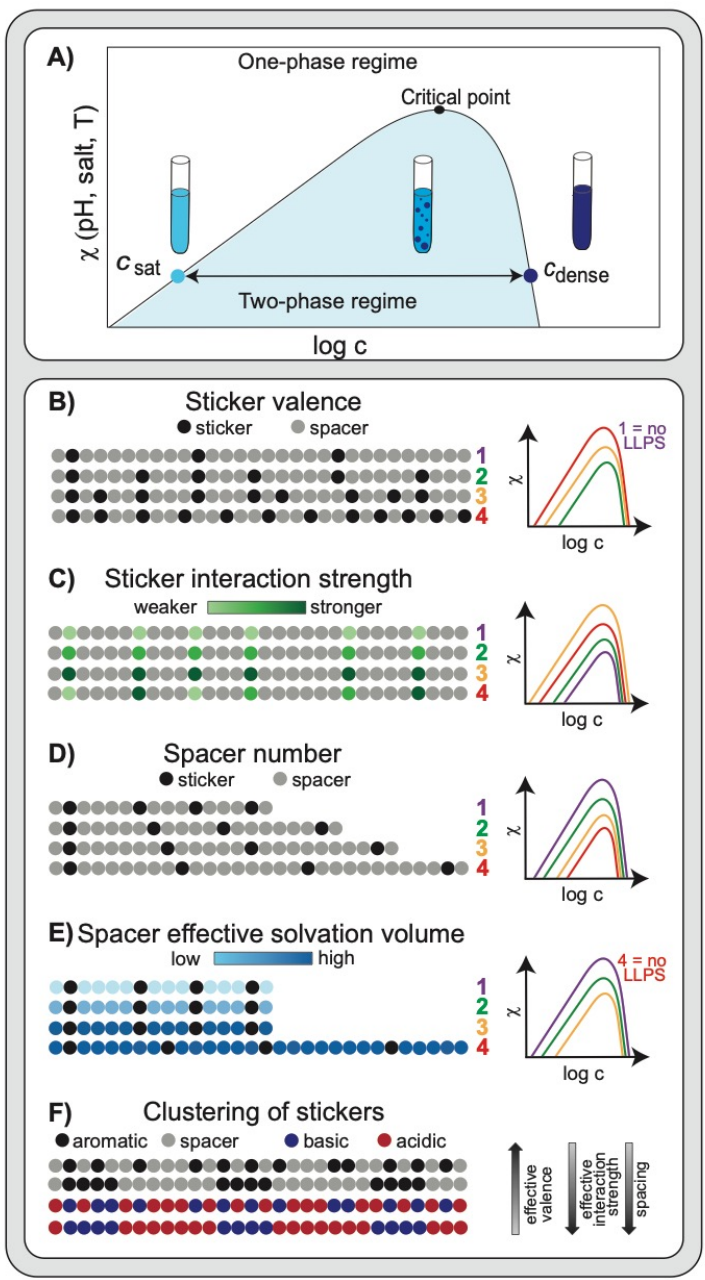

**Figure 2. Sticker valence, interaction strength and patterning, and spacer properties determine biomolecule phase behavior. (A)** The schematic phase diagram shows the coexistence curve for a

biomolecule as a function of interaction potential $\chi$, which can, e.g., be modulated by temperature, pH and salt concentration. As the coexistence curve is crossed and the biomolecule enters the two-phase regime, the biomolecule undergoes LLPS and forms a dilute and a dense phase, whose concentrations are given by the left and right arms of the coexistence curve, i.e., $c_{sat}$ and $c_{dense}$. The zenith of the curve is the critical point beyond which no phase separation occurs. Increasing the driving forces for phase separation will result in a widening of the two-phase regime and an increase in the position of the critical point. **(B)** The sticker valence determines the driving force for LLPS. Increasing the sticker valence increases the driving force for phase separation; too few stickers results in the absence of a two-phase regime. **(C)** Increasing the sticker interaction strength increases the driving force for LLPS. **(D)** Increasing the number of spacers between the stickers decreases the driving force for phase separation by reducing the cooperativity of sticker-sticker interactions. **(E)** The effective solvation volume of spacers determines whether the formation of three-dimensional networks is coupled to a density transition, i.e., LLPS. In the case of well-solvated spacers, stickers can mediate the formation of a system-spanning network even in the absence of LLPS. **(F)** Clustering of stickers alters multiple properties of the biomolecule simultaneously: It decreases the effective valence of the system, increases the effective interaction strength of the stickers and increases the spacing between the stickers. In the case of aromatic stickers, clustering can promote amorphous aggregation over LLPS.

Other PTMs that have been implicated in modulating phase separation include citrullination of arginine and acetylation of lysine both of which remove positive charge. Citrullination of arginine has been shown to reduce recruitment of FUS to stress granules [70], likely because arginine-mediated interactions are removed. Acetylation of lysine disrupts phase separation of tau by disrupting the electrostatic interactions which support its phase separation [71]. In the case of poly-ADP-ribosylation, the extending chain of poly-ADP-ribose (PAR) may introduce additional multivalence, thus increasing the driving force for phase separation [72]. In each of these examples the stickers-and-spacers framework can help explain and predict the effect of the PTM.

**Conclusion**
In summary, the stickers-and-spacers or similar frameworks provide simple but powerful schemes to conceptually understand the driving force for phase separation of IDRs and which elements in a sequence are most important for their phase behavior. Such models have helped to quantify the effects of increasing and decreasing sticker valence, predict the impact of altering sticker strength and patterning, and understand the effects of spacer solvation. They have also proven useful for understanding the effects of PTMs on the phase behavior of IDRs. In addition, they help us understand which sequences do not have relevant driving forces for phase separation because they miss adhesive elements and are highly soluble due to their preferred interaction with solvent. Further developments of the stickers-and-spacers framework should continue to provide insights into the effects of sequence alterations, helping to provide a mechanistic understanding of disease mutations. Challenges for the future include modeling the effects of conformational differences between the dilute and dense phase and their effects on phase behavior. The stickers-and-spacers framework promises to advance our ability to predict phase behavior from sequence alone, improving our understanding of biomolecular condensate formation and providing a strong physical basis for the understanding of a multitude of biological processes.

**Acknowledgments**
We thank members of the Mittag lab and our colleagues Rohit Pappu, Richard Kriwacki, Paul Taylor, Kresten Lindorff-Larsen and Alex Holehouse for fruitful discussions. T.M. acknowledges funding by NIH grant R01GM112846, by the St. Jude Children's Research Hospital Research Collaborative on Membrane-less Organelles in Health and Disease, and by the American Lebanese Syrian Associated Charities. The content is solely the responsibility of the authors and does not necessarily represent the official views of the National Institutes of Health.

**Conflict of Interest**

T.M. is a consultant for Faze Medicines. This affiliation has not influenced the scientific content of this review.

# 39. Lin Y, Currie SL, Rosen MK: **Intrinsically disordered sequences enable modulation of protein phase separation through distributed tyrosine motifs**. *J Biol Chem* 2017, **292**:19110-19120.

40. Murthy AC, Dignon GL, Kan Y, Zerze GH, Parekh SH, Mittal J, Fawzi NL: **Molecular interactions underlying liquid-liquid phase separation of the FUS low-complexity domain**. *Nat Struct Mol Biol* 2019, **26**:637-648.

41. Dignon GL, Best RB, Mittal J: **Biomolecular phase separation: From molecular driving forces to macroscopic properties**. *Annu Rev Phys Chem* 2020, **71**:53-75.

42. Brangwynne Clifford P, Tompa P, Pappu Rohit V: **Polymer physics of intracellular phase transitions**. *Nature Physics* 2015, **11**:899-904.

43. Hughes MP, Sawaya MR, Boyer DR, Goldschmidt L, Rodriguez JA, Cascio D, Chong L, Gonen T, Eisenberg DS: **Atomic structures of low-complexity protein segments reveal kinked beta sheets that assemble networks**. *Science* 2018, **359**:698-701.

44. Luo F, Gui X, Zhou H, Gu J, Li Y, Liu X, Zhao M, Li D, Li X, Liu C: **Atomic structures of FUS LC domain segments reveal bases for reversible amyloid fibril formation**. *Nat Struct Mol Biol* 2018, **25**:341-346.

45. Gui X, Luo F, Li Y, Zhou H, Qin Z, Liu Z, Gu J, Xie M, Zhao K, Dai B, et al.: **Structural basis for reversible amyloids of hnRNPA1 elucidates their role in stress granule assembly**. *Nat Commun* 2019, **10**:2006.

46. Peran I, Mittag T: **Molecular structure in biomolecular condensates**. *Curr Opin Struct Biol* 2020, **60**:17-26.

47. Semenov A, Rubinstein M: **Thermoreversible Gelation in Solutions of Associative Polymers. 1.Statics**. *Macromolecules* 1998, **31**:1373-1385.

48. Rubinstein M, Dobrynin A: **Solutions of associative polymers.** *Trends in Polymer Science* 1997, **5**:181-186.

49. Choi JM, Holehouse AS, Pappu RV: **Physical principles underlying the complex biology of intracellular phase transitions**. *Annu Rev Biophys* 2020, 10.1146/annurev-biophys-121219-081629.

** Extensive review of the stickers-and-spacers formalism and other coarse-grained models for protein phase behavior. It discusses strategies for differentiating between stickers and spacers, understanding effective *vs* apparent valence and the influence of spacer properties on phase separation. It discusses the influence of the lifetimes of sticker-sticker crosslinks on protein dynamics in condensates and material properties.

50. Choi JM, Hyman AA, Pappu RV: **Generalized models for bond percolation transitions of associative polymers**. *arXiv* 2020, **2004.03278**.

51. Choi JM, Dar F, Pappu RV: **LASSI: A lattice model for simulating phase transitions of multivalent proteins**. *PLoS Comput Biol* 2019, **15**:e1007028.

52. Kim YC, Hummer G: **Coarse-grained models for simulations of multiprotein complexes: Application to ubiquitin binding**. *J Mol Biol* 2008, **375**:1416-1433.

53. Dignon GL, Zheng W, Kim YC, Best RB, Mittal J: **Sequence determinants of protein phase behavior from a coarse-grained model**. *PLoS Comput Biol* 2018, **14**:e1005941.

** This report demonstrates that simple amino acid-specific interaction potentials can be used to model and predict phase behavior via coarse-grained off-lattice simulations. It also demonstrates
14

the potential of such models towards generating understanding of the effects of PTMs on phase behavior.